\documentclass{article}
\usepackage{ijcai20}

\usepackage{times}

\usepackage{soul}
\usepackage{url}
\usepackage[hidelinks]{hyperref}
\usepackage[utf8]{inputenc}
\usepackage[small]{caption}
\usepackage{graphicx}
\usepackage{amsmath}
\usepackage{booktabs}
\usepackage{bm}
\usepackage{amsfonts}
\usepackage{caption}
\usepackage{subcaption}
\usepackage{multirow}
\title{Interpretable Multimodal Learning for Intelligent Regulation in Online Payment Systems}

\author{
Shuoyao Wang$^1$\footnote{Work done while Shuoyao was a senior researcher with Financial Technology Group, Tencent, China.}\footnote{
Corresponding author.}\and
Diwei Zhu$^2$\and
\affiliations
$^1$College of Electronic and Information Engineering, Shenzhen University, China\\
$^2$Department of Information Engineering, The Chinese University of Hong Kong, Hong Kong\\
\emails
w.shuoy@gmail.com, zd115@ie.cuhk.edu.hk
}

\begin{document}

\maketitle

\begin{abstract}
With the explosive growth of transaction activities in online payment systems, effective and real-time regulation becomes a critical problem for payment service providers. Thanks to the rapid development of artificial intelligence (AI), AI-enable regulation emerges as a promising solution. 
One main challenge of the 
AI-enabled regulation is how to  utilize multimedia information, i.e., multimodal signals, in Financial Technology (FinTech).
Inspired by the attention mechanism in nature language processing, we propose a novel cross-modal and intra-modal attention network (CIAN) to investigate the relation between the text and transaction.
More specifically, we integrate the text and transaction information to enhance the text-trade joint-embedding learning, which clusters positive pairs and push negative  pairs  away  from  each  other.
Another challenge of intelligent regulation is the interpretability of complicated machine learning models. To sustain the requirements of financial regulation, 
we design a CIAN-Explainer to interpret how the attention mechanism interacts the original features, which is formulated as a low-rank matrix approximation problem.
With the real datasets from the largest online payment system,  WeChat Pay of Tencent, we conduct experiments to validate the practical application value of CIAN, where our method outperforms the state-of-the-art methods. 
\end{abstract}

\section{Introduction}
As a new term in the financial industry, FinTech has become a popular term that describes novel technologies adopted by the financial service institutions. At the juncture of these phenomena, the risk of online transaction and shopping becomes increasingly prominent. Establishing a reliable intelligent regulation infrastructure, e.g., to detect whether the transaction flow of a merchant is beyond  the scope of its licensed business, is essential to accommodating more FinTech startups and e-commerce. 
Thanks to the advance of personal computers, smart phones, and internet, the quantity of digitized multimedia contents has increased dramatically \cite{WinNT}. 
Consequently,   
despite of the efforts utilizing 
subtle feature engineering and classifiers for AI-enabled regulation \cite{cao2019titant}, the efficient utilization of digitized multimedia contents is still a core research challenge for intelligent regulation.

Inspired by the recent advances in deep learning,
multimodal machine learning has been widely employed to interpret such multimedia information during the past few years. These techniques can be roughly classified into four categories: multimdedia speech recognition \cite{8585066}, multimdedia content analysis (e.g., automatic shot boundary detection \cite{lienhart1998comparison}, video summarization \cite{zhang2016video}), human behavior understanding (e.g., emotion recognition \cite{chu2016selective}), and multimodal matching (e.g., visual question-answering \cite{antol2015vqa}, image-text matching \cite{wang2019position}).

In this paper, we focus on the general line regulation problem, i.e., to detect whether the transaction flow of a merchant is beyond  the scope of its licensed business, which can be formulated as a multimodal matching problem.
In the past few years, many research efforts have been devoted to multimodal matching. For instance, \cite{ijcai2019-299} and  \cite{ijcai2019-503} concatenated features from different modalities via a fusion layer  for sentiment analysis and crowdfunding success prediction, respectively. 
Inspired by the great success of attention in nature language processing (NLP), many studies have validated the attention is helpful to model a more reliable relationship between image and text
\cite{ijcai2019-568,ijcai2019-431,wang2019position}.  
Most recently, a popular framework to model the multimodal relationship is the two-branch embedding network, where one branch encodes the first modality information and another modules the other one
\cite{wang2018learning,gu2018look,wang2019position}. 
\begin{figure*}[!h]
\centering
\includegraphics[width=0.9\textwidth]{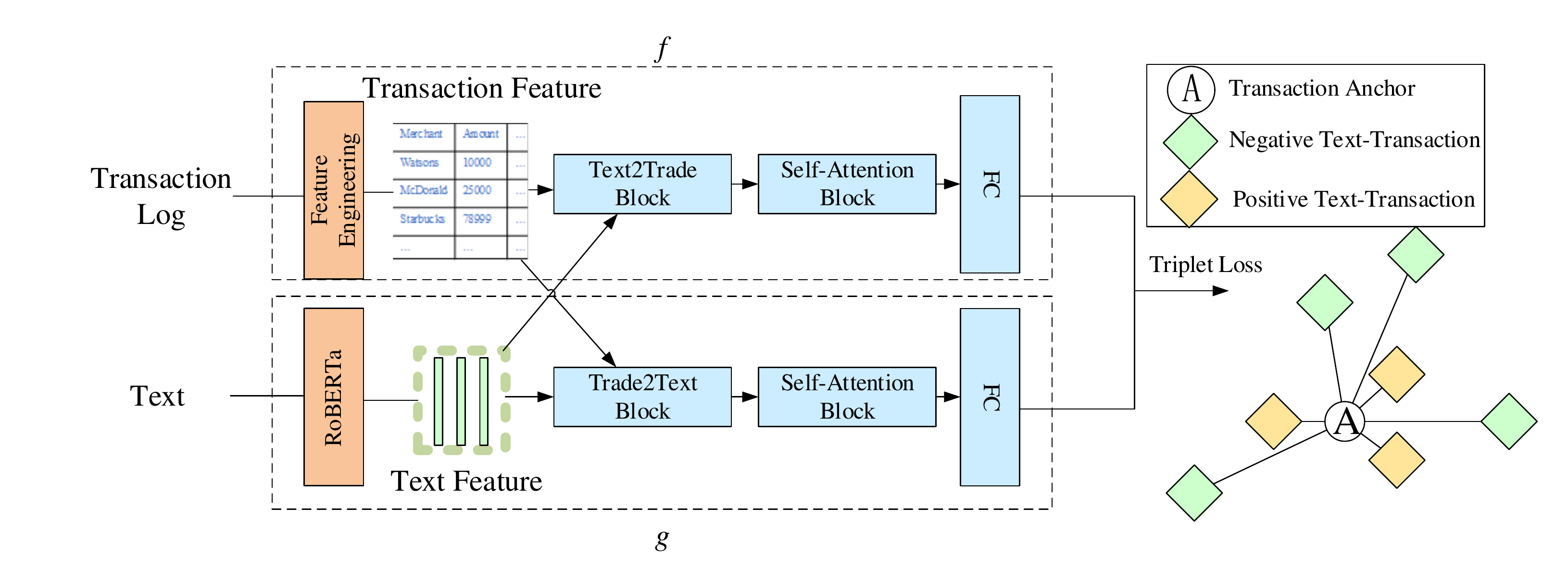}
	\caption{CIAN consists of two modules: (i) the feature extraction module which extracts text and transaction features using their corresponding backbone architectures; (ii) the  cross-modal and inter-modal attention projection that match the
feature distributions originating from the same identity. The network is trained by the incremental-margin triplet loss.
}
	\label{fig:network}
\end{figure*}
It is desired if the cross-modal and intra-modal relationships could be jointly investigated in a unified two-branch embedding network for AI-enabled regulation problem.
Unfortunately, the existing multimodal learning methods and attention mechanisms mostly focus on the textual and visual information. To the best of our knowledge, transaction flow information as the most important modality in FinTech has not been studied yet. 

In particular, the intra-modality relations within text and transaction flow are complementary to the inter-modality relations between text and transaction flow, which were mostly ignored by existing mechanisms.
For example, if the description of a merchant indicates it is a restaurant,  the meal-time transaction flows express the main semantics of the merchant with high probability, while the gender of the customers may not be that important.
Moreover, on one hand, fully exploiting interaction relations leads to complex models and thus becomes hard to explain. On the other hand,  “need to explain” is already enforced in many industries, especially in FinTech.
To tackle the gap,  we propose a novel cross-modal and intra-modal  attention network (CIAN) to investigate the relation between the textual and transaction flow views. Moreover, we also propose a CIAN-Explainer and formulate it as an optimization problem, i.e., a low-rank matrix approximation problem. Consequently, CIAN-Explainer identifies a small subset of transaction flow features that have a crucial role in transaction-text matching. 
The major contributions are summarized as follows:
\begin{itemize}
    \item To the best of our knowledge, this paper is among the first to propose an attention mechanism to  text and transaction flow  for AI-enabled regulation.
    \item We propose both intra-modal and cross-modal attention to capture the correspondences inside the financial behavior,  and between the financial behavior and text description.
\item  We also investigate a novel CIAN-Explainer, which provides a variety of benefits, from the ability to visualize semantically relevant features to interpretability, to saving a huge amount of human-resources for manually auditing.
\item We validate the performance and evaluate the application value on a practical e-payment dataset, which is with heavy noise. The results show that the performance of the proposed method is significantly better than all the benchmark methods with regard to different evaluation metrics.  
\end{itemize}

\section{Problem and Methodology}
In this section, we first define the problem accurately and then elaborate the details of CIAN: a cross-modal matching approach  that learns to match the feature representations from the two modalities in order to perform both
text-to-transaction and transaction-to-text retrieval. 

\subsection{Motivation}
In a multimedia e-commerce system (e.g., WeChat Pay), there is a set of resident merchants $\mathcal{V}$. Each merchant $i \in \mathcal{V}$ is composed of its transaction flow logs $T_i$ (transaction information) and descriptions $D_i$ (text information). Without confusion, we would use the term of transaction and text interchangeably in this paper. 
In the e-commerce system, each merchant is only allowed to carry out the designated business. 
In order to avoid potential harm to consumer and small business borrowers, merchants should be appropriately regulated at their licensed business scope.
As a result, if the descriptions of a merchant are conflict with its transaction logs, the merchant will be suspended or limited.
However, the merchants could fake the descriptions to match the designated business, which are at high-risk of regulation.
To tackle this problem, we propose CIAN to investigate the transaction-text matching problem and figure out the dismatching merchants, which could significantly enhance the FinTech regulation.

\subsection{Joint Feature Learning}
In our multimodal matching problem, one of the main objectives is to learn discrimiative transaction and text feature representations that accurately retrieve transaction/text from text/transaction. 
Fig.~\ref{fig:network} shows the the training procedure.
Specifically, the CIAN framework consists of a transaction encoder and a text encoder, which are described in detail below.
\subsubsection{Transaction Encoder}
\begin{figure}[!htp]    
	 \centering\includegraphics[width=0.38\textwidth]{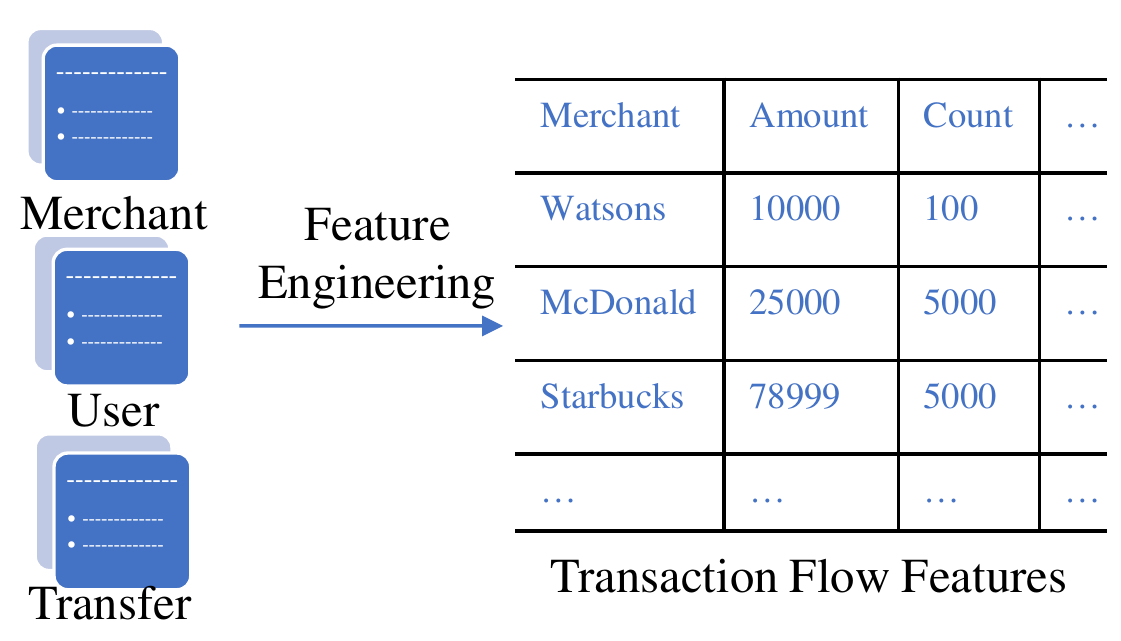}
	\caption{Transaction flow features are extracted from user profile, merchant profile, transfer environment, and etc.}
	\label{fig:transaction_encoder}
\end{figure}
As the problem of AI-enabled regulation is vital to a financial business, efforts have been spent for years, where about 415 features are carefully engineered in WeChat Pay system. We call such features as
transaction flow features. Specifically, given transaction flow logs $T_i \in \mathcal{T}$, we obtain the transaction features  $\bm{t}_i=[t_{i,1},t_{i,2},...,t_{i,415}] \in \mathbb{R}^{415}$ (Fig.~\ref{fig:transaction_encoder}).

\subsubsection{Text Encoder}
Similarly, the corresponding description of  a merchant is also represented by features through a text encoder. Thanks to the recent advance in NLP, we adopt RoBERTa \cite{liu2019roberta} as the text encoder, where all sentences are padded and truncated to the same length 128. Given the description of merchant $D_i \in \mathcal{D}$, we obtain the text features $\bm{d}_i=[d_{i,1},d_{i,2},...,d_{i,768}] \in \mathbb{R}^{768}$.
\subsubsection{Objective}
We denote the transaction and text encoder as function $f: \mathcal{T} \to \mathbb{R}^{1024}$ and $g: \mathcal{D} \to \mathbb{R}^{1024}$, which map transaction and text to vectors with the same dimension $1024$, respectively (Fig.\ref{fig:network}).
For a transaction-text pair $(T_i,D_i)$ of merchant $i$, the similarity $S_i$ is measured by cosine similarity:
\begin{equation}
    S_i= \left<\frac{f(T_i)}{||f(T_i)||_2},\frac{g(D_i)}{||g(D_i)||_2}\right>: \mathcal{T} \times  \mathcal{D} \to \mathbb{R}.
\end{equation}
According to the similarity, the network is trained by the incremental-margin triplet loss \cite{zhang2019learning}, which clusters positive pairs and push negative pairs away from each other. There are mainly two kinds of blocks inside $f$ and $g$: cross-modal attention blocks and intra-modal attention blocks, which will be introduced in the next section.
\subsection{Attention Mechanism}
Intuitively, the transaction behavior and description of a merchant are highly correlated and coupled. For example, if the description of a merchant indicate it is a restaurant,  the meal-time transaction flows  may express the main semantics of the merchant with higher probability, while the genders of the customers may not be significant. However, if the description is highly related to makeup, the genders of the customers become very essential. 
To tackle this issue, we propose the cross-modal attention blocks and inter-modal attention blocks as shown in Fig.~\ref{fig:network}. In the remaining of this section, we explain the underlying attention mechanisms employed at the blocks.
\subsubsection{Cross-modal Attention} 
\begin{figure}[htp!]
\centering
\begin{subfigure}{.5\linewidth}
\centering
\includegraphics[width=\linewidth]{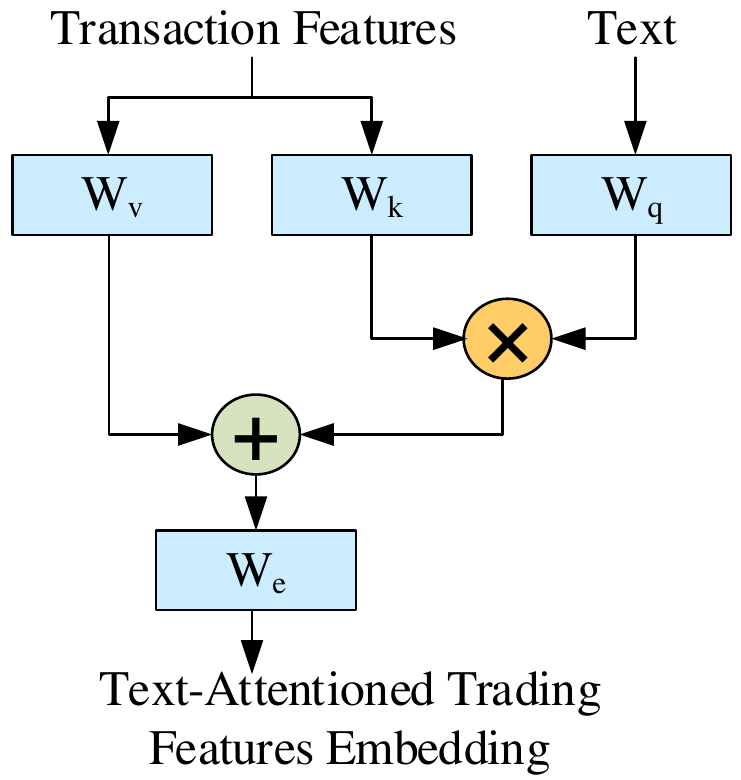}
\caption{}\label{AT14}
\end{subfigure}%
\begin{subfigure}{.5\linewidth}
\vspace{-0.2cm}
\centering
\includegraphics[width=\linewidth]{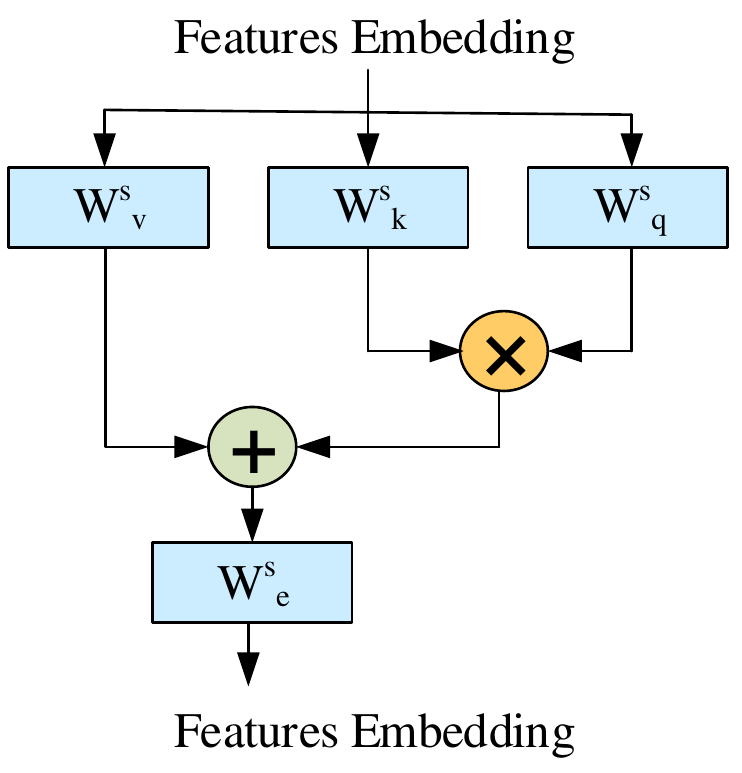}
\caption{}\label{NO14}
\end{subfigure}
\caption{\small(a) Text to Transaction Attention Block. (b) Intra-modal Attention Block. $\times$ denotes the inner product operation; $+$ denotes the element-wise multiplication operation.}\label{fig:tion.test}
\end{figure} 
The cross-modal attention blocks  learn to capture the interaction between different modalities. 
In particular, we adopt the text to transaction attention block (Fig.~\ref{AT14}) as an example to illustrate how the cross-modal attention works. Then, the transaction to text attention block could be easily understood by swapping the transaction features and text features. 
Each merchant transaction and text features are first  transformed into value, key, and query  features following \cite{vaswani2017attention,gao2019dynamic}, where the transformed features are denoted as $\bm{v}_i$, $\bm{k}_i$, $\bm{q}_{i} \in \mathbb{R}^{\text{dim}}$. Notation $\text{dim}$ represents the common dimension of transformed features from both modalities\footnote{In our simulation, we set $\text{dim}=1024$.}.
Given merchant $i$, the block first captures the value $\bm{v}_i$ and key $\bm{k}_i$ from the transaction features:
\begin{equation}
    \begin{aligned}
        \bm{v}_{i}&=\bm{W}_v \times \bm{t}_i+\bm{b}_v, \\
        \bm{k}_{i}&=\bm{W}_k \times \bm{t}_i+\bm{b}_k,
    \end{aligned}
\end{equation}
and the query $\bm{Q}_i$ from the text features:
\begin{equation}
        \bm{q}_{i}=\bm{W}_q \times \bm{d}_i+\bm{b}_q, 
\end{equation}
where $\bm{W}_v$, $\bm{W}_k$, $\bm{W}_q$, and $\bm{b}_v$, $\bm{b}_k$, $\bm{b}_q$ are the weights and bias, respectively. 
The block then packs $\bm{k}_{i}$ and $\bm{q}_{i}$ together and calculate their attention to $\bm{v}_{i}$:
\begin{equation}
        \bm{a}_i=\text{softmax}\left(\frac{\bm{k}_i\bm{q}_i^T}{\sqrt{\text{dim}}}\right),
\end{equation}
where $\bm{a}_i \in \mathbb{R}^{\text{dim}}$ is the set of attention weights. 
The dot product values are proportional to the dimension of the common feature space. Consequently, the product is normalized by $\sqrt{\text{dim}}$. The softmax non-linearity function is applied row-wisely.
Notice that the package operation could be Gaussian $e^{\bm{k}_i\bm{q}_i^T}$,  Kernel Gaussian $e^{\phi (\bm{k}_i) \phi (\bm{q}_i)^T}$, dot product $\bm{k}_i\bm{q}_i^T$, and etc. Interestingly, we will show by experiments that CIAN is not sensitive to the choices and thus choose the dot product operation for its simplicity and interpretability.
With the attention weights, the block outputs the embedding features simply by element-wise multiplication as shown in Fig.~\ref{AT14}:
\begin{equation}
    \bm{o}_i=\bm{a}_i \circ \bm{v}_i,
\end{equation}
where $\bm{o}_i$ is the cross-modal attention output of merchant $i$.
\subsubsection{Intra-modal Attention}
On top of the cross-modal attention, we  also observe the relationships within each modality are complementary to the cross-modal relations. For instance,  we should pay more attention to the gender related features when we figure out the merchant is related to makeup. Meanwhile, if the gender related features are essential, we should also focus on the repurchase rate due to the difference of purchasing habits among different genders.

Therefore, we also design intra-modal attention block in $f$ and $g$. The implementation of intra-modal attention block is illustrated at Fig.\ref{NO14}, which differs from the cross-modal attention block only on the input and thus we omit the details due to the page limitation.
\section{CIAN-Explainer}
To sustain the requirements of financial regulation, we should not only detect whether the transaction flow of a merchant is beyond the scope of its licensed business but also figure out the reason of the dis-match. Consequently, we also design a CIAN-Explainer to interpret how the attention mechanism interact the original features.
\begin{figure}[!htp]    
	 \centering\includegraphics[width=0.40\textwidth]{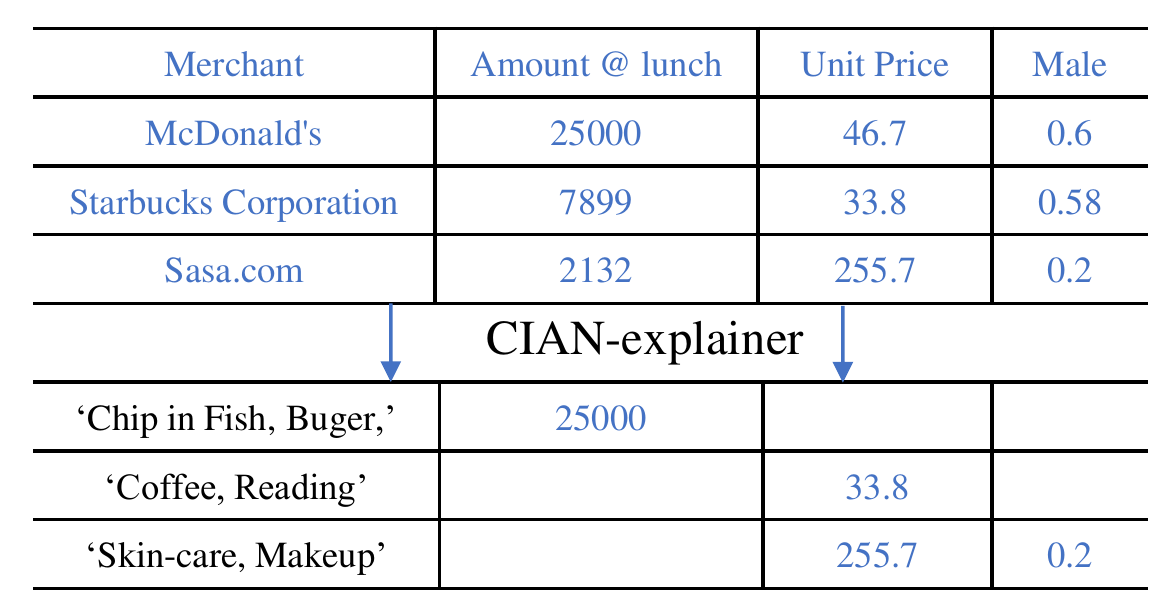}
	\caption{CIAN-Explainer specifies an explanation as a rich subset of the entire transaction flow information.}
	\label{fig:explain}
\end{figure}
In this section, we propose a CIAN-Explainer, an approach for explaining matching made by CIAN. The proposed CIAN-Explainer takes a trained CIAN  and its match results as input, and returns an explanation in the form of a small subset of transaction flow features that are most crucial for the match (Fig.~\ref{fig:explain}). 
CIAN-Explainer specifies an explanation as a rich subset of the entire transaction flow information, such that the subset minimizes
the distance between the approximate embedding and real embedding of CIAN match.
Based on this, we formulate CIAN-Explainer as a low-rank approximation problem:
\begin{subequations}\label{eq:pf}
\begin{align}
\underset{\hat{a}}{\text{min}}&~~ ||W_v\left(t_i \circ \hat{a}\right)+b_v-\bm{a}_i \circ \bm{v}_i||_2^2\\
\text{s.t.}&~~ ||\hat{a}||_1 \leq 10, \\
&~~||\hat{a}||_2=1.
\end{align}
\end{subequations}
Unfortunately, the constraint  $||\hat{a}||_1 \leq 10$ describes a non-convex set, which makes Problem~(6) hard to solve. Thanks to the duality principle, we could re-formulate CIAN-Explainer as the dual problem of Problem~(6):
\begin{subequations}\label{eq:pf}
\begin{align}
\underset{\hat{a}}{\text{min}}&~~ ||W_v\left(t_i \circ \hat{a}\right)+b_v-\bm{a}_i \circ \bm{v}_i||_2^2+\lambda||\hat{a}||\\
\text{s.t.}&~~||\hat{a}||_2=1,
\end{align}
\end{subequations}
where $\lambda$ is the Lagrange multiplier and the problem could be solved by alternating direction method of multipliers (ADMM).
As a result, the explainer learns a real-valued feature mask $\hat{a}$ which selects the important transaction flow features and masks out unimportant ones.
In our experiments, we evaluate CIAN-Explainer and the results show that CIAN-Explainer provides consistent and concise explanations of CIAN match. The visualization of CIAN-Explainer will be shown in Section~4.3.

\section{Experiments}
In order to demonstrate the effectiveness of our proposed methods, we conduct CIAN on a practical e-payment dataset: WeChat Pay. We systematically make comparisons with several latest start-of-the-art methods and thoroughly investigate the performance of the proposed CIAN framework. As for the performance measure criterion for transaction-text matching, we apply the commonly used precise, recall, and F1-Score as the performance evaluation metrics. 
\subsection{Implement Details}
\subsubsection{WeChat Pay Dataset}
For a merchant, its transaction flow records make up a basic data identification, and the remarks and comments are regarded as the description of this transaction flow. 
From the real-world datasets in WeChat Pay, we extract two sub-datasets for the evaluation.
In the first sub-dataset, namely 31-D dataset, we collect 50,000 merchants from 2019/07/01 to 2019/07/31 for training, 1,000 merchants from 2019/08/01 to 2019/08/31 for validating, and 1,000 merchants from 2019/07/01 to 2019/07/31 for testing, from WeChat Pay system. 
In the second dataset, namely 7-D dataset, we collect 50,000 merchants from 2019/07/01 to 2019/07/07 for training, 1,000 merchants from 2019/08/01 to 2019/08/07 for validating, and 1,000 merchants  from 2019/07/01 to 2019/07/07 for testing.
Through the hardest negative sampler \cite{hermans2017defense}, there are total 500,0000 training pairs, 10,000 pairs for validating, and 10,000 for testing.
In our datasets, we define the positive pairs as the merchants' transaction flow stats and their corresponding descriptions, and the negative pairs as the merchants' transaction flow stats and the descriptions of the merchants who belong to other categories.
\begin{table*}[!h]
\centering
\label{my-label2}
\scriptsize
\begin{tabular}{@{}c|c|c|c|c|c|c|c|c@{}}
\toprule
\multirow{2}{*}{Model} & \multirow{2}{*}{Component} &\multirow{2}{*}{Output Layer} & \multicolumn{3}{c|}{7-D Dataset}&\multicolumn{3}{c}{31-D Dataset}\\ \cmidrule{4-9}
& & & Precise & Recall &F1-Score& Precise & Recall &F1-Score\\ \midrule
\multirow{2}{*}{FC} & \multirow{2}{*}{NAN} &Fusion Layer+Classification\cite{ijcai2019-299} &53.0 & 57.8 & 55.3& 55.3  & 60.3 & 57.7  \\ \cmidrule{3-9}
 &  &Triplet Loss\cite{wang2018learning}& 50.0 & 50.1 &50.1 &51.2&50.3&50.7\\ \midrule
\multirow{2}{*}{Bi-Linear Attention} & \multirow{2}{*}{NAN} & Fusion Layer+Classification \cite{kim2018bilinear}&57.6 & 58.1 & 57.8 &60.0&60.2& 60.1\\ \cmidrule{3-9}
& &Triplet Loss&65.8 & 64.1 & 64.9 &69.1 &66.7&67.9\\ \midrule
\multirow{6}{*}{CIAN} & \multirow{2}{*}{Intra-modal Attention}&Fusion Layer+Classification&57.8&59.1&58.4& 59.8&61.3&60.5\\ \cmidrule{3-9}
& &Triplet Loss&69.0 &67.0 & 68.0& 71.9 & 70.0&70.9\\ \cmidrule{2-9}
& \multirow{2}{*}{Cross-modal Attention}&Fusion Layer+Classification&59.4 &58.8 &59.1& 62.2&61.9&62.0\\ \cmidrule{3-9}
& &Triplet Loss&77.0&68.5&72.5& 80.9 & \textbf{71.4}&75.9\\ \cmidrule{2-9}
& \multirow{2}{*}{Both} &Fusion Layer+Classification&59.0&58.9&58.9& 61.7&61.5&61.6\\ \cmidrule{3-9}
& &Triplet Loss&79.5 & 69.9 & 74.4&\textbf{85.9} & \textbf{71.4}&\textbf{78.0}\\ \bottomrule
\end{tabular}
\caption{Comparisons of transaction-text matching on WeChat Pay dataset with the competing methods.}
\end{table*}

\subsubsection{Training details} 
All of our experiments are conducted on  a machine with an Intel
Xeon E5-2630 CPU, two NVIDIA GTX 1080 Ti GPUs, and 64GB
RAM. 
The text embedding is achieved with the pre-trained RoBERTa \cite{liu2019roberta}.\footnote{\url{https://github.com/brightmart/roberta_zh}} 
After text embedding and feature engineering, the dimension of transaction and text information are $415$ and $768$, respectively.
For information interaction, the transaction and text information are projected into the same dimension, which is fixed as $1024$ in our experiment. 
The networks are constructed using Pytorch for computational speed boost. 
The model parameters are initialized with kaiming normal initializer and Adam optimization algorithm is used to train the overall network. 
Moreover, we set the batch size to 256, the initial learning rate to $0.01$ and the regularizer parameter as $0.01$ to prevent over-fitting.
\subsection{Embedding Visualization}
\begin{figure}[htp!]
\centering
\includegraphics[width=\linewidth]{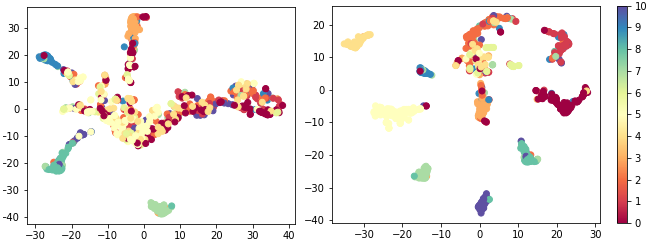}
\caption{\small Two-dimensional visualization of transaction features before and after cross-modal embedding. Colors represent :  Restaurant, Shopping Mall, Travel, Express, Education, Technology, Healthy, Jewellery, Monopoly, Clothes, and Makeup from 0 to 10.  }\label{fig:clu}
\end{figure}

We use t-SNE \cite{maaten2008visualizing} to visualize our transaction features before and after cross-modal embedding, coloring each merchant according to its category in Fig.\ref{fig:clu}. 

In the two-dimensional visualization, the left figure shows that the transaction features before cross-modal embedding are vaguely clustered but the boundaries merge together and some categories are diffuse. Meanwhile, in the right figure, the transaction features after cross-modal embedding are less noisy and clearly clustered corresponding to the merchant category. The above observations corroborate our motivation of combining intra-modal and cross-modal attention mechanism in our CIAN design.
Through cross-model embedding, more attentions are paid to the information that helps identify the scope of business.

\subsection{Performance Evaluation}
\subsubsection{Comparison versus Different Package Operations}
We observe through our experiments that CIAN is not sensitive to the choices of the package operation in Eq.~(4) and thus choose the dot product operation for its simplicity and interpretability.
To validate the observation, we exhibit the comparison among different types of the package operation in Table 2.
We can see that the Gaussian, Kernel Gaussian, and Dot product versions achieve the similar performance, up to some random variations (77.5 to 78.3). Due to the simplicity, we adopt dot product in the proposed CIAN framework. Thanks to the simplicity, we proposed the CIAN-Explainer in section~III, which provides inspiring insights and save a huge amount of human-resources for manually auditing.
\begin{table}[htp]
\centering
\label{my-label1}
\scriptsize
\begin{tabular}{@{}c|c|c|c|c|c|c@{}}
\toprule
\multirow{2}{*}{Operation} & \multicolumn{3}{c|}{7-D Dataset}&\multicolumn{3}{c}{31-D Dataset}\\ \cmidrule{2-7}
& Precise & Recall &F1-Score& Precise & Recall &F1-Score\\ \midrule
Gaussian& \textbf{80.5} &  \textbf{70.3} & \textbf{75.1} & 85.5 &  70.9 & 77.5 \\ \midrule
Kernel Gaussian &80.1 &69.8& 74.6 &\textbf{75.0} &70.5& \textbf{78.3} \\ \midrule
Dot product& 79.5 & 69.9 & 74.4& 85.9 & \textbf{71.4} & 78.0\\ \bottomrule
\end{tabular}
\caption{Comparisons of the choices of package operation in Eq.~(4).}
\end{table}
\begin{figure}[!htp]    
	 \centering\includegraphics[width=0.42\textwidth]{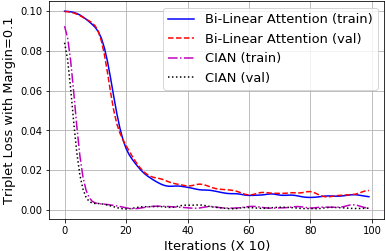}
	\caption{Curves of the training procedure for CIAN  \emph{vs}  the Bi-linear Attention baseline with 31-D dataset. We show the training and validation triplet loss, i.e., the objective function during the training procedure. The matching precise and recall results are in Table 1.}
	\label{fig:curve}
\end{figure}

\subsubsection{Comparison with Competing Methods}
\begin{figure*}[!h]
\begin{subfigure}{.66\linewidth}
\centering
\includegraphics[width=\linewidth]{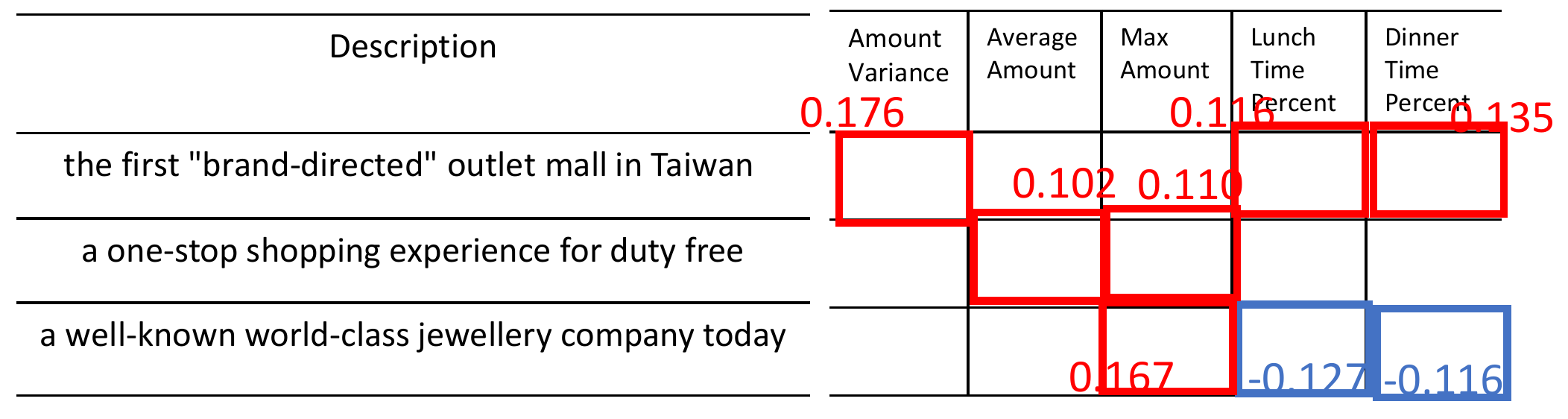}
\caption{Examples of the relationship between the description and CIAN-Explainer.}\label{fig:exp}
\end{subfigure}%
\begin{subfigure}{.34\linewidth}
\centering
\includegraphics[width=0.75\linewidth]{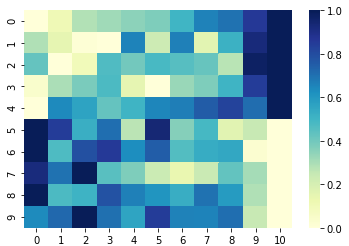}
\caption{Projected attention of different categories.}\label{fig:cluster}
\end{subfigure}
\caption{Visualization of CIAN-Explainer. In (b), the columns represent 11 categories: Restaurant,  Shopping Mall, Travel,Express, Education, Technology, Healthy, Jewellery, Monopoly, Clothes, and Makeup from $0$ to $10$; the rows stand for 10 selected features, which is not elaborated due to the non-disclosure agreement.}\label{fig:test}
\end{figure*}
Fig.~\ref{fig:curve} shows the curves of the training procedure of our proposed CIAN framework  \emph{vs}  the Bi-Linear Attention baseline \cite{kim2018bilinear} with 31-D dataset. Under both 7-D dataset and 31-D dataset,  our proposed CIAN model is consistently better than the Bi-Linear Attention baseline throughout the training procedure, in both training and validation error.
Moreover, the convergence rate of CIAN is much faster than the baseline and the converged performance of CIAN also outperforms the baseline.

Table~1 exhibits the performances of all methods on WeChat Pay dataset. The major findings from the experimental results can be summarized as follows:
\begin{itemize}
    \item Our model outperforms all the baselines, which indicates our model adopts a more principled way to leverage intra-modal information and cross-modal relations for improving match performance. The performance of all the methods under 31-D dataset is notably better than the one under 7-D dataset. Our model achieves 85.9\% precise, which is a remarkable performance under the real-world data with randomness and noise. 
    \item Among these baselines, we can find that the overall performance order is as follows: (attention+triplet loss) based methods, (fully-connected layers+triplet loss) based methods, (attention+fusion layer+classification) based methods, (fully-connected layers+fusion layer+classification) based methods. It indicates that the better performances can be achieved through attention mechanism and two-branch embedding structure with triplet loss.
    \item Comparing the only cross-modal attention network and only intra-modal attention network, we can find that the cross-modal attention network outperforms the intra-modal attention one, which further demonstrates the importance of  capturing the correspondences inside the trading behavior and between the trading behavior and the meaning of the sentences.
\end{itemize}

\subsection{Explainer Visualization}
To sustain the requirements of financial regulation, we design a CIAN-Explainer to interpret how the attention mechanism interact the original features in Section~III.
In Fig.~\ref{fig:test}, we visualize the CIAN-Explainer to analyze transaction and text matching for WeChat Pay.
An exemplary visualization result is shown in Fig.~\ref{fig:exp}.  There are two major observations:
\begin{itemize}
    \item The correspondences between the text and  transaction are satisfied, most of the phrases can attend to
their related transaction features, like the phrases ``outlet mall'', ``duty free'', ``Jeweller'', and so on.
\item The sign of the  attention indicates the positive or negative text-transaction correlation. The absolute value of attention could filtered out unrelated information flows, like the average transaction amount of a outlet mall.
\end{itemize}

Intuitively, the merchants with similar text should focus on relative features. In order to see if the learned attention preserves the similarity, we cluster the merchants with their categories and then plot the average attentions for selected 10 features in Fig.~\ref{fig:cluster}. 
We observe that the closer categories share higher attention similarity in most cases. For example, merchants of Restaurant and Shopping Mall, i.e., column 0 and 1 in Fig.~\ref{fig:cluster}, both pay more attentions on the features related to the transaction time distribution and pay less attentions on the features related to the gender distribution. However, merchants of Clothes and Makeup, i.e., column 9 and 10 in Fig.~\ref{fig:cluster}, focus on the features related to the repurchase rate and gender. As a consequence, CIAN-Explainer could not only visualize relevant features, but also save a huge amount of human-resources for manually auditing, which is quiet important in financial business.

\section{Conclusion}
In this paper, we develop a novel cross-modal and intra-modal attention network (CIAN) for the transaction-text matching task. Our proposed CIAN  framework can be adapted to any cross-modal matching problem, e.g., image-text, image-transaction, transaction-text. We show the CIAN framework alternatively passes information within and across transaction and text  based on the proposed attention mechanism. To sustain the requirements of financial regulation, we also design a CIAN-Explainer to interpret how the attention mechanism interacts the original features. Furthermore,  we collect a dataset from a practical online payment (WeChat Pay) and make the first attempt to evaluate the application value of our transaction-text model, which provides solid improvement over baselines. We hope CIAN as well as CIAN-Explainer provides a big insight for both the AI-enabled FinTech and interpretable multimodal learning. 
\section*{Acknowledgments}
This work is supported in part by General Research Funding (project number 14200315) established by Hong Kong Research Grant Council. 
\newpage
\bibliographystyle{named}
\bibliography{ijcai20}
\end{document}